\documentclass[nofootinbib,preprint,aps,draft,pre,superscriptaddress,citeautoscript,floatfix]{revtex4-1}

\usepackage{graphics}
\usepackage{graphicx}
\usepackage{mathrsfs}
\usepackage{color}
\usepackage{bm}

\begin{document}

\title{Vapor-Induced Motion of Liquid Droplets on an Inert Substrate}
\author{Xingkun Man}
\email{Email: manxk@buaa.edu.cn}
\affiliation{Center of Soft Matter Physics and its Applications, Beihang University, Beijing 100191, China}
\affiliation{School of Physics and Nuclear Energy Engineering, Beihang University, Beijing 100191, China}
\author{Masao Doi}
\email{Email: masao.doi@buaa.edu.cn}
\affiliation{Center of Soft Matter Physics and its Applications, Beihang University, Beijing 100191, China}
\affiliation{School of Physics and Nuclear Energy Engineering, Beihang University, Beijing 100191, China}

\begin{abstract}

Evaporating droplets are known to show complex motion that has conventionally been explained by the
Marangoni effect (flow induced by the gradient of surface tension). Here, we show that the droplet
motion can be induced even in the absence of the Marangoni effect due to the gradient of
evaporation rate. We derive an equation for the velocity of a droplet
subject to non-uniform evaporation rate and non-uniform
surface tension placed on an inert substrate where
the wettability  is uniform and unchanged.
The equation explains the previously observed attraction-repulsion-chasing
behaviors of evaporating droplets.

\end{abstract}

\maketitle

Evaporating droplets show complex motion that
has attracted abiding interest in scientific
research since the early work of Leidenfrost who observed chaotic motion of water droplets on hot skillet~\cite{Leiden56}.
Equally complex motion of droplets (attraction,
repulsion, chasing etc) has been reported for volatile droplets
slowly evaporating at room temperature~\cite{Bang38,Cottington64,
Carles89,Bahadur09,Sellier13,Cira15}, but,
apart from the conjecture that the phenomena is
caused by the Marangoni effect (flow induced by the gradient
of surface tension), no quantitative theory has been
given.

Cira et al~\cite{Cira15} reported that evaporating droplets show
complex motion even if they are placed on an inert substrate where the wettability is uniform and unchanged.
Figure~1 schematically shows their results for the case of pure liquid droplets.
Here, droplets made of pure water
(W) and polypropylene glycol (PG) are evaporating on a
solid substrate. The left droplet is mobile and the right droplet is
pinned. They reported that W moves away
from PG (Fig.~1b), but PG moves
toward W (Fig.~1c), (hence they chase each
other when both droplets are mobile). On the other hand,
same droplet pair (W-W and PG-PG) always attract
each other (Fig.~1a and d).

Such motion of evaporating droplets has conventionally been
explained by the vapor-mediated Marangoni
effect~\cite{Cottington64,Carles89,Bahadur09,Sellier13,Cira15}.
Liquid vapor evaporating from one droplet
condenses in other droplets and changes the local
composition of the droplet (see Fig.~1e).
Since the vapor density is not uniform  (the vapor density
is high near the droplet and decreases with the distance
$r$ from the droplet),
this creates a non-uniform surface tension in the droplet, and
causes a Marangoni flow leading to the droplet motion.

The Marangoni effect can explain the observed motion of two
droplets made of different kinds of liquids (W-PG), but cannot explain
the motion of two droplets of same pure liquid (W-W or PG-PG)
since there should be no Marangoni effect in this case.

The effect of non-uniform vapor density is not limited to
the Marangoni effect. If the vapor density is not uniform,
the evaporation rate of the droplet becomes non-uniform
\cite{Cira15,Deegan00,Kobay10,Pradhan15}.
This effect is important for the the
droplets of same pure liquid as it is the only mechanism for the
vapor-mediated interaction between them.

Non-uniform evaporation of the liquid induces internal
fluid flow, and thus deforms the droplet.
Daniel et al~\cite{Daniel01} have pointed out
that such shape deformation actuates the motion of the droplet
due to the gradient of the Laplace pressure, but this effect
(called Capillary effect by Brenner et al~\cite{Wasan01})
has not been analysed theoretically.

\begin{figure}
\begin{center}
\includegraphics[bb=0 0 420 250, scale=0.95,draft=false]{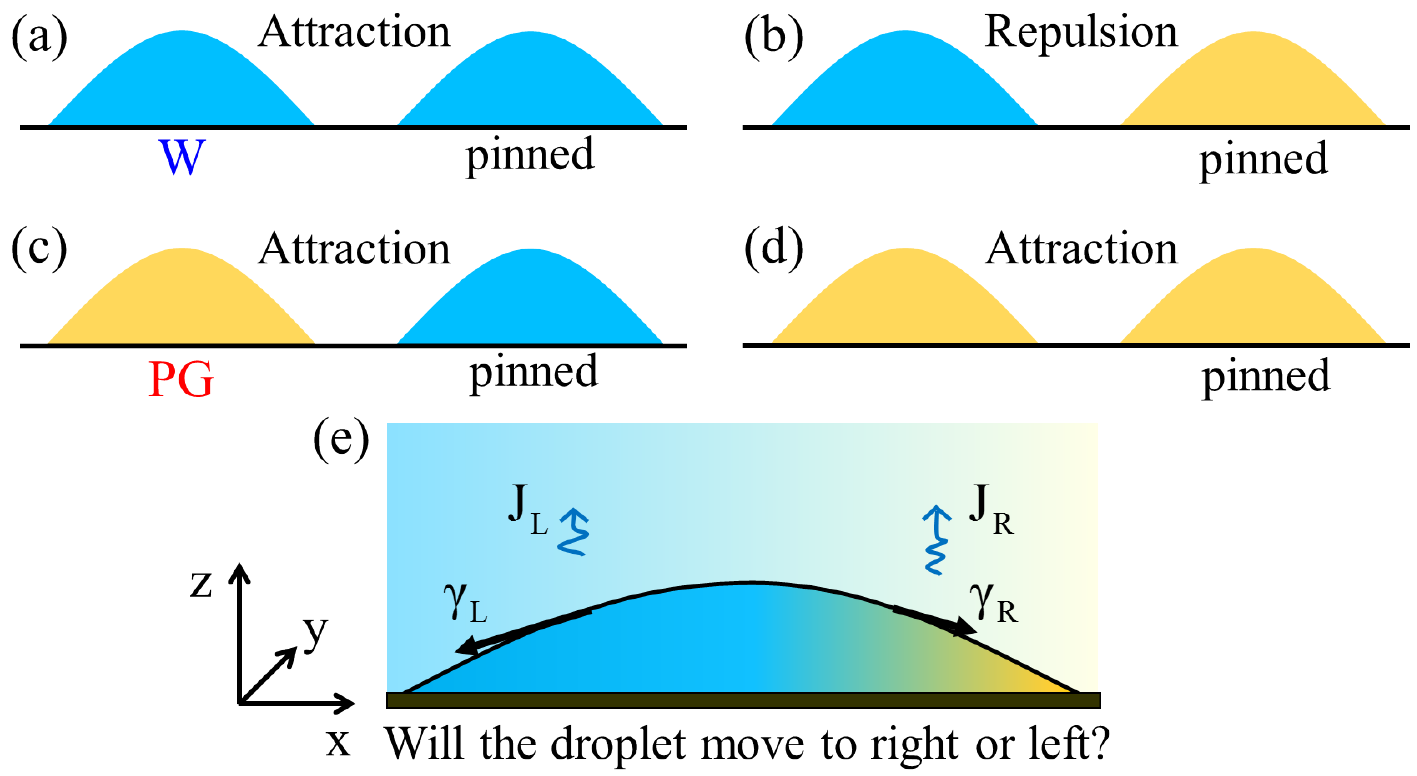}
\caption{(a)-(d) Schematic pictures of the directional motions of
two neighbouring droplets reported by
Cira et al, Nature 519, 446 (2015). Blue droplet denotes a pure water
(W) and yellow droplet denote a pure propylene glycol (PG).
In all figures, the right droplet is pinned, while the left one is free.
(a) Two W droplets attract each other; (b) A pinned PG droplet
repels a free W droplet; (c) A pinned W droplet attracts
a free PG droplet; (d) Two PG droplets attract each
other. (e) The problem discussed in this article: how a droplet which
has asymmetrical liquid/vapor surface tension moves when it
is placed in a non-uniform vapor environment.}
\label{fig1}
\end{center}
\end{figure}

In the following we consider the motion of a
droplet placed on a substrate and in non-uniform vapor
density (see Fig.~1e).  To avoid complications, here we assume
that the substrate is inert: the wettability of the substrate is uniform
and is not affected by the vapor.  If the wettability changes, it will
creates the motion of a droplet.  This problem
has been extensively studied both experimentally~\cite{Chaudhury92,
Brzoska93,Gallardo99, Ichimura00,Style13}
and theoretically~\cite{Brochard89,Xu12},
and can be easily included in the present theory.

Due to the non-uniform vapor density, the evaporation rate
$J$ and the liquid/vapor surface tension $\gamma$ are not uniform.
Here we assume that the gradients of $J$ and $\gamma$
are small and have $x$-component only:
\begin{eqnarray}
J&=&J_0+\frac{\partial J}{\partial x}\left[x-x_c(t)\right]\label{f1},\\
\gamma&=&\gamma_0+\frac{\partial \gamma}{\partial x}\left[x-x_c(t)\right]\label{f2},
\end{eqnarray}
where $x_c(t)$ is the center of the contact line of the droplet,
and $J_0$ and $\gamma_0$ are the mean evaporation rate
and the surface tension.

To determine the motion of this droplet, we use the Onsager
principle~\cite{Doi13,Doi15}. This principle is
equivalent to the variational principle known in Stokesian
hydrodynamics for moving boundary problem:
the motion of moving boundaries is determined by
the condition that the Rayleighian $\Re=\Phi+\dot{F}$
be minimum with respect to the boundary velocity,
where $\Phi$ is the energy dissipation function
(the half of the energy dissipation rate created in the
fluids by the bounday motion), and $\dot{F}$
is the change rate of the free energy. The principle has
been applied for the droplet motion by gravity~\cite{Xianmin16} and by
evaporation~\cite{Man16}.

To use this principle, we assume that the surface
profile of the droplet is given by (in a cylindrical coordinate)
\begin{equation}\label{f3}
h(r,\varphi,t)=H(t)\left[1-\frac{r^2}{R(t)^2}\right]
                   \left[1+\alpha(t)rcos\varphi\right],
\end{equation}
where $H(t)$ and $R(t)$ are the height and the radius of the droplet,
and $\alpha(t)$ is a parameter describing the shape deformation of the
droplet.  The state of the droplet is specified by four
parameters $H(t)$, $R(t)$, $\alpha(t)$ and $x_c(t)$. We shall
determine the time evolution of these parameters by the
Onsager principle.

We assume that the droplet is nearly flat [$R(t)\gg H(t)$]
and use the lubrication approximation to calculate
the dissipation function:
\begin{equation}\label{f4}
\Phi(\bm{v}_f)=\frac{1}{2}\int^{R}_{0}\int^{2\pi}_{0}
                     \frac{3\eta}{h} \bm{v}^2_f rdr d\varphi,
\end{equation}
where $\eta$ is the viscosity of the fluid and
$\bm{v}_f(r,\varphi)$ is the height averaged velocity of the fluid
at point $(r,\varphi)$. Therefore, the model is valid when the
droplet contact angle and the fluid Reynolds number are small, i.e. $\theta \ll 1$ and $Re\ll 1$. 
The The velocity is determined by
the minimum of $\Phi(\bm{v}_f)$ under constraint that
$\bm{v}_f$ satisfies the conservation equation
$\dot h =- \nabla\cdot(\bm{v}_f h ) - J $.

The calculation of $\bm{v}_f$ becomes simple in the special
case that the droplet remains undeformed [i.e, $\alpha(t)$
remains to be zero]. In this case, the $x$
component of  $\bm{v}_f$ is given by
\begin{equation}\label{f5}
v^{x}_{f}=v_0+\dot x_c+\frac{R^2}{2H}\frac{\partial J}{\partial x},
\end{equation}
where $v_0$ stands for the velocity induced by $J_0$ and is
independent on $\dot x_c$. Inserting this expression
into Eq.~(4), $\Phi$ is calculated as
\begin{equation}\label{f6}
\Phi=\Phi_0+\frac{3\pi C\eta R^2}{H}\dot x^2_c+\frac{3\pi C\eta R^4}{2 H^2}\frac{\partial J}{\partial x}\dot x_c,
\end{equation}
where $\Phi_0$ represents the term arising from $v_0$ and is
independent of $\dot x_c$, and $C$ is a dimensionless
constant given by $C=\ln \left(R/2\epsilon\right)$
where $\epsilon$ is the molecular cutoff length that is introduced
to remove the divergence in the energy dissipation
at the contact line.

The rate of the free energy change $\dot F$ has two parts:
the capillary effect, $\dot F_C$, and the Marangoni
effect, $\dot F_M$. $\dot F_C$ is
independent of $\dot x_c$,
while $\dot F_M$ is written as
\begin{eqnarray}\label{f7}
\dot F_M &=& -\int^{R}_{0}\int^{2\pi}_{0}\bm{v}_s\cdot \nabla \gamma rdrd\varphi \nonumber\\
&=&-\frac{3\pi R^2}{2}\frac{\partial \gamma}{\partial x} \dot x_c,
\end{eqnarray}
where $\bm{v}_s$ is the fluid velocity on the liquid/vapor interface,
which is related to $\bm{v}_f$ by
$\bm{v}_s=3\bm{v}_f/2+(\partial \gamma/\partial x)h/4\eta$.

By minimizing the Rayleighian $\Re=\Phi+\dot F$ with respect to $\dot x_c$, we obtain the velocity of the droplet induced
by the gradients of surface tension and evaporation rate:
\begin{equation}\label{f8}
\dot x_c=\frac{R\theta}{4\eta C}\frac{\partial \gamma}{\partial x}-\frac{R}{\theta}\frac{\partial J}{\partial x},
\end{equation}
where $\theta(t)=2H(t)/R(t)$ is the contact angle (see the Supplemental Material for the full derivation).

\begin{figure}
\begin{center}
\includegraphics[bb=0 0 520 290, scale=0.88,draft=false]{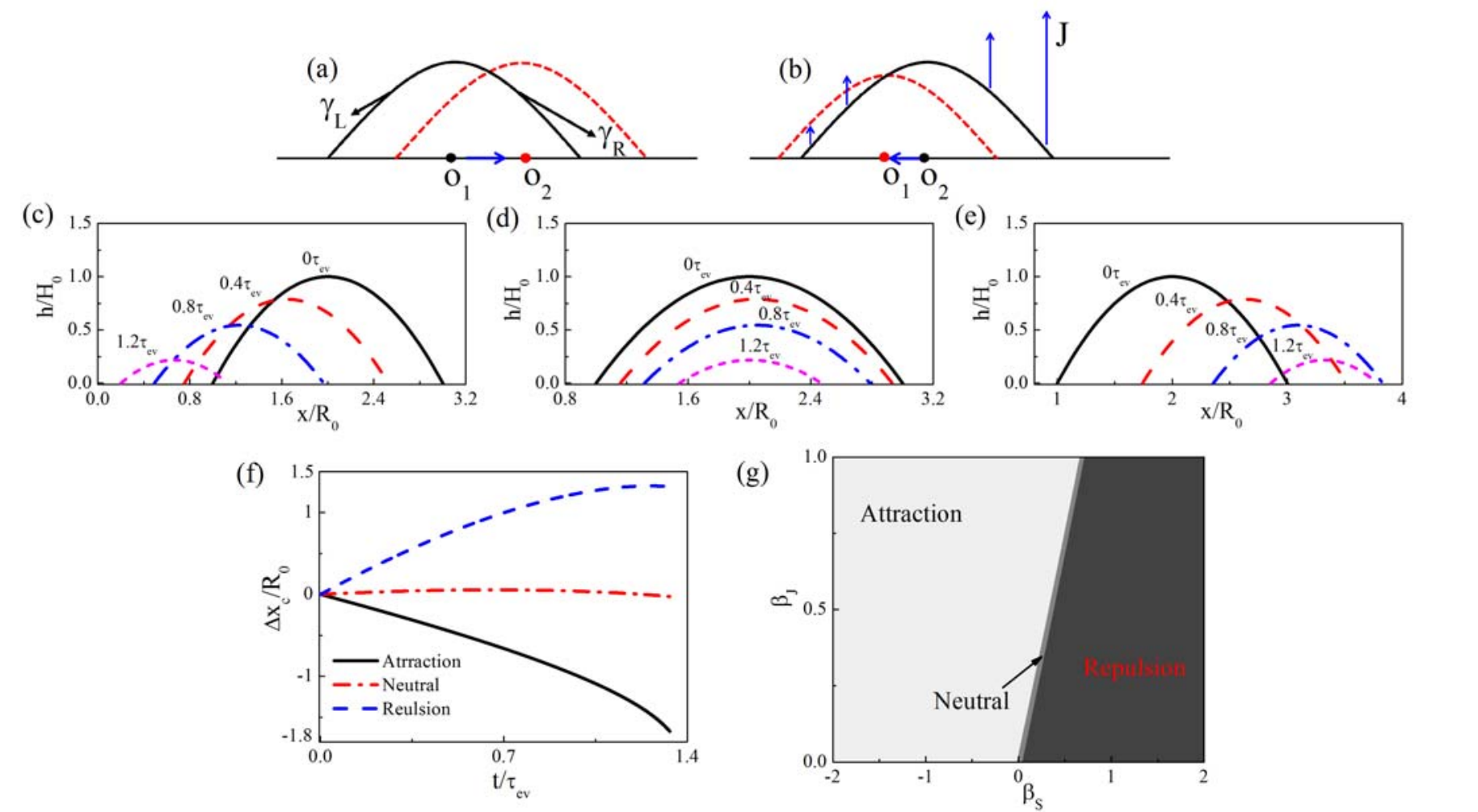}
\caption{(a-b) Schematic pictures of theoretical results for the motion of liquid
droplet induced by non-uniform surface tension (a)
and by non-uniform evaporation (b).
(c-e) Calculated time evolution of the droplet interface between the liquid and vapor, $h(x,y=0,t)/H_0$.
Here $H_0$ and $R_0$ are the initial value of $H(t)$ and $R(t)$,
and the dimensionless surface tension gradient $\beta_S$ and the
evaporation rate gradient $\beta_J$ (defined in the text)
are given by (c): $\beta_S=-0.2$, $\beta_J=1.0$, (d): $\beta_S=0.4$,
$\beta_J=0.5$, and (e): $\beta_S=2.0$, $\beta_J=1.0$.
(f) Calculated displacement of the droplet.
(g) Phase diagram for the direction of the droplet motion.}
\label{fig2}
\end{center}
\end{figure}

Equation~(\ref{f8}), indicates that non-uniform
surface tension $\partial \gamma/\partial x$ moves
the droplet from low surface tension side to high
surface tension side.  This conclusion is consistent with the
understanding that droplet moves in the same direction
as the Marangoni flow which is in the direction toward the high surface tension side (see Fig.~2a).

Equation (\ref{f8}) also indicates that the non-uniform
evaporation rate $\partial J/\partial x$
moves the droplet from high evaporation side to low
evaporation side (evaporation effect).
This result can be understood by the minimum energy
dissipation principle in Stokesian hydrodynamics. Consider the
case of $\partial J/\partial x>0$ (see Fig.~2b), the liquid on the right side of
the droplet evaporates faster than that on the left side.
To maintain the symmetric
parabolic shape given by Eq.~(\ref{f3}) with
$\alpha=0$, fluid flow is created in the droplet. If the contact line is
fixed, this fluid flow is from left to right. However, if the contact line is mobile, the motion of the droplet takes place from right to left to
reduce the energy dissipation associated with this fluid flow.

For a pair of same pure liquid droplets, $\partial \gamma/\partial x$
is zero, while $\partial J/\partial x $ is non zero.  Since
the evaporation rate is lower in the middle of the pair than at the edge,
$\partial J/\partial x $ is negative for the left droplet and positive for the
right one. Accordingly, they always approach each other.
This explains the experimentally observed motions of pair W or PG droplets~\cite{Cira15}
which have not been explained so far.

For a pair of droplets made of different pure liquids,
evaporation rate will not be
affected ($\partial J/\partial x \simeq 0$),
but the surface tension will be
changed locally due to the condensation of the vapor of other
droplets. Since the surface tension of a liquid usually increases
when vapor of other liquid having higher surface tension condenses,
the droplt moves towards the droplet having higher surface
tension. Similarly, the droplet having higher surface tension
moves away from the droplet having lower surface tension. Thus the
lower surface tension droplet chases larger surface tension droplet.

\begin{figure}
\begin{center}
\includegraphics[bb=0 0 500 220, scale=0.9,draft=false]{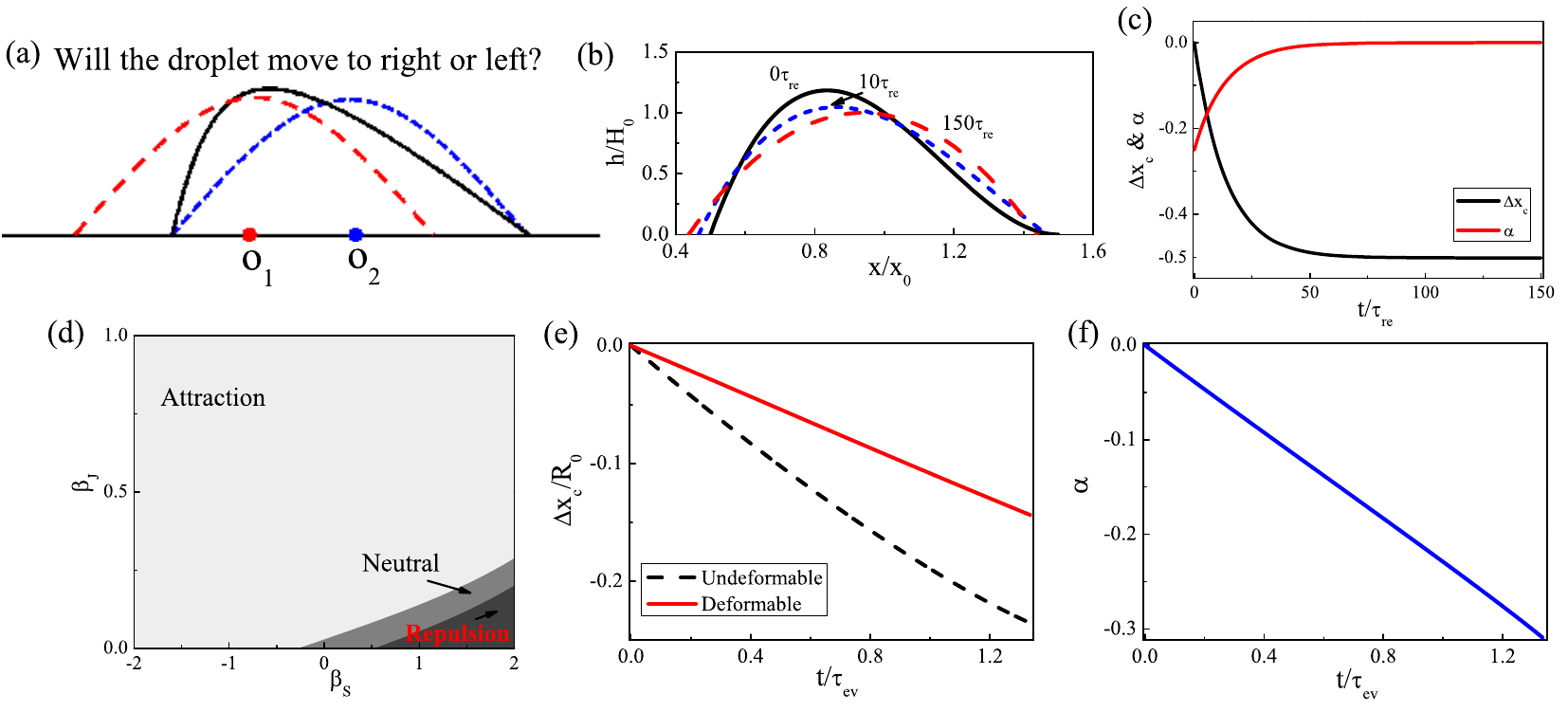}
\caption{
(a) An unaddressed problem: will a droplet with asymmetric
initial shape move to the right or left?
(b) Calculated evolution of the profile of the droplet interface between the liquid and vapor, $h(x,y=0,t)/H_0$, for $\alpha(0)=-0.25$, $J(t)=0$ and $\beta_S=0$.
(c) Displacement of the center of the droplet and the relaxation of the shape deformation shown in (b). Here the time is in the units of $\tau_{\rm{re}}$ (the characteristic relaxation time for a deformed droplet, defined by $\tau_{\rm{re}}=\eta V^{1/3}_0/\gamma_0 \theta^3_e$, where $\theta_e$ is the equilibrium contact angle).
(d) Phase diagram for the direction of motion of deformable droplet in
$(\beta_S, \beta_J)$ space. All parameters are the same as in Fig.~2(g), but the droplet is deformable.
(e) Comparison of $\Delta x_c/R_0$ between deformable and undeformable droplets
for $\beta_J=0.2$ and $\beta_S=-0.1$.
(f) Deformation parameter $\alpha(t)$ of the deformable droplet
for the situation shown in (e) is plotted against $t$.}
\label{fig3}
\end{center}
\end{figure}

When the droplets are made of solutions, both effects
of $\partial J/\partial x $ and $\partial \gamma/\partial x$ become
important.  To study the direction of the droplet motion,
we solved the time evolution equations obtained by the
Onsager principle, and the results are shown in Fig.~2c-e.
Here $\partial \gamma/\partial x$ and $\partial J/\partial x $
are assumed to be constant, and are represented by
dimensionless quantities $\beta_S$ and $\beta_J$ which are defined by
$\beta_S =\partial \gamma/\gamma_0\partial x $ and
$\beta_J = \tau_{\rm{ev}} \partial J/\partial x$, where
$\tau_{\rm{ev}}$ is the characteristic evaporation time
defined by $\tau_{\rm{ev}}=V_0/|\dot V_0|$ ($V_0$ and $R_0$ are the initial size and radius of the droplet).

Figure~2f shows the evolution of the displacement of the droplet
$\Delta x_c=x_c(t)-x_c(0)$ for the three cases of (c), (d) and (e).
It is seen that when the mobile droplet is attracted to the fixed droplet,
the speed accelerates. This is
consistent with the experimental results~\cite{Cira15}.

Figure~2g shows the phase diagram for the direction of motion in the
parameter space of ($\beta_S$, $\beta_J$). Since the velocity $\dot x_c$
changes in time, we define the direction of the droplet motion according to the final moving displacement $\Delta x_f$: Attraction is
for $\Delta x_f<-0.001R_0$, Neutral is for $-0.001R_0\leq\Delta x_f\leq0.001R_0$, and Repulsion is for $\Delta x_f>0.001R_0$.

We have discussed droplets which keep parabolic
shape during evaporation [i.e., $\alpha(t)$ is assumed to be zero].
We now consider droplets that is deformable [i.e., $\alpha(t)$ can be
non-zero].

We first consider a problem how a droplet that is
initially deformed as shown by the solid black line in
Fig.~3a [i.e., $\alpha(0)<0$] will move as the
shape relaxes to the equilibrium one.
This problem is non-trivial. If we focus on the
force acting on the contact line, the excess force acting on the left
contact line $\gamma_{SV}-\gamma_{SL}-\gamma\cos \theta_L$
($\gamma_{SV}$ and $\gamma_{SL}$ are the substrate/vapor and
substrate/liquid surface tension) is larger than the force acting on the right.
Hence the droplet is expected to move to the
left~\cite{Daniel01}. On the other hand,
the gradient of the pressure difference in the droplet
produces a capillary flow from left to right~\cite{Wasan01}.
These two forces acts in opposite direction. To our knowledge,
no existing theory predicts the direction of the droplet motion
in this situation.

We can answer this question by Onsager principle,
which gives the following time evolution equations when the effects of
shape deformation, surface tension gradient and
evaporation rate gradient are included (see the Supplemental Material)
\begin{eqnarray}
\dot x_c
&=&
\frac{R\theta^3}{3\eta C}\frac{\partial\gamma}{\partial x}-\frac{R}{\theta}\frac{\partial J}{\partial x}+\frac{\gamma_0R\theta^3}{3\eta C}\alpha\label{f9},\\
\dot \alpha
&=&
\frac{2\theta}{\eta RC}\left(1-\frac{4\theta^2}{3}\right)\frac{\partial\gamma}{\partial x}-\frac{8\gamma_0\theta^3}{3\eta RC}\alpha\label{f10}.
\end{eqnarray}
Figure~3b and c show an example of the solution of the
above equation for the case of no evaporation [$J(t)=0$] and
no Marangoni effect ($\partial \gamma/\partial x=0$) with
an initially asymetric shape [$\alpha(0)=-0.25$].
It is seen that the droplet quickly relaxes to its equilibrium
shape and stops. As the shape relaxation takes place,
the droplet moves from small contact angle side to the
large contact angle side: the direction opposite to the flow
induced by the Laplace pressure. Again, this result can be
explained in terms of the minimum-energy dissipation principle:
if the contact line is pinned, the fluid flows from the large contact
angle side to the small contact angle side; but if the contact line is mobile,
the center of the droplet move from the small contact angle side to
the large contact angle side in order to reduce the energy dissipation
associated with the fluid flow.

Figure~3d shows a phase diagram of the direction of the
droplet motion in ($\beta_S$, $\beta_J$) plane for deformable
droplet. Compared with Fig.~2g, it is seen that the area of
attraction becomes larger for deformable droplet than for
undeformable droplet: the Marangoni effect on the motion
becomes weaker.

Figure~3e shows the trajectory of $\Delta x_c$ for
both deformable and undeformable droplets. The undeformable droplet moves faster
than the deformable droplet. Figure~3f is the evolution
of $\alpha(t)$, which clearly indicates that the motion
is accompanied by the shape deformation.
Although a negative $\alpha$ enhances the approaching
velocity of droplets, the velocity of deformable
droplet is smaller than the undeformable one.

This can be understood as follows. Equation~(\ref{f10})
indicates that a part of the free energy is
used to deform the shape of a droplet, while it is
fully used to move the droplet that is undeformable
(or more rigid). Therefore, for deformable droplets, the
migration velocity becomes smaller, and larger value of
$\beta_S$ is needed to change the direction of the droplet motion.

The theory presented so far can be used to simulate the
motion of many droplets. Figure~4 shows the motion of two droplets
made of different kinds of liquids placed on an inert solid substrate.
The gradients of the evaporation rate and surface tension of the droplets
are assumed to be constant and are set
as in the caption.  This parameter set gives the
positive migration velocity for both droplet with the left
larger than the right, and causes a chasing motion.

\begin{figure}
\begin{center}
\includegraphics[bb=0 0 180 235, scale=1.2,draft=false]{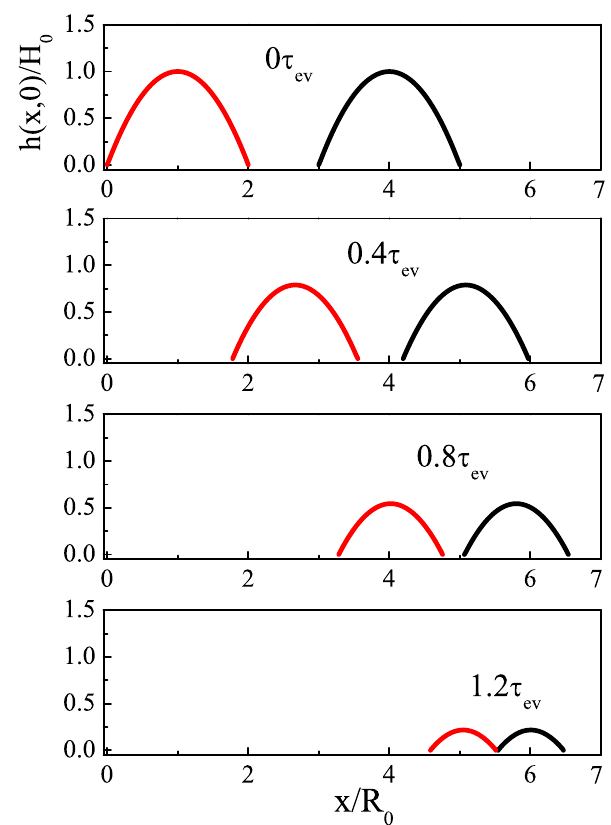}
\caption{Simulation of the motion of two evaporating
droplets. Parameters used in this calculation are:
$\beta^L_S=3.5$, $\beta^R_S=3.0$, $\beta^L_J=-1.0$ and $\beta^R_J=1.0$. Chasing is observed.}
\label{fig4}
\end{center}
\end{figure}

In the present theory, we have assumed that the surface property of
the substrate is unaffected by vapor and remains uniform.
If a gradient is created for the surface property,
an additional force is created which drives the droplet from
high surface energy region to low surface energy region.
This problem has been theoretically discussed
by Brochard et al~\cite{Brochard89} and Xu et al~\cite{Xu12}
and they predicted
\begin{equation}\label{f11}
\dot x_c  =  \alpha_v\frac{H}{\eta C}\frac{\partial \gamma_{SL}}{\partial x},
\end{equation}
where $\alpha_v$ is a positive constant that depends
on the shape of the droplet.
The same equation can be derived from the
Onsager principle (see Supplemental Material).
Equation (\ref{f11}) can be added to
the evolution Eq.~(\ref{f9}) if there is a gradient in
the surface energy of the substrate.

To summarize, we have presented a theory for the motion of
evaporating droplets on an inert substrate that remains uniform.
The theory explains that
(a) in a same pure liquid droplet pair, they always attract each other,
(b) in a different pure liquid pair, the lower surface tension droplet
chases larger surface tension droplet, and (c) in a solution droplet pair, they attract, repel or chase each other depending on the surface tension and the evaporation rate. The theory can also be used to simulate the complex motion of many droplets evaporating on a substrate.

We thank T. A. Witten and D. Andelman for useful discussions. This work
was supported in part by Grant No. 21404003 and 21434001
of the National Natural Science Foundation of China (NSFC), and
the joint NSFC-ISF Research Program, jointly funded by the
NSFC under Grant No. 51561145002 and the Israel
Science Foundation (ISF) under Grant No. 885/15.



\begin{thebibliography}{99}
\bibitem{Leiden56} J. G. Leidenfrost, \emph{De aquae communis nonnullis qualitatibus tractatus} (Duisburg, 1756); transl. C. Wares, Int. J. Heat Mass Transfer $\bf9$, 1153 (1966).

\bibitem{Bang38} D. H. Bangham and Z. Saweris, Trans. Faraday Soc. $\bf34$, 554 (1938).

\bibitem{Cottington64} R. L. Cottington, C. M. Murphy, and C. R. Singleterry, Adv. Chem. $\bf43$, 341 (1964).

\bibitem{Carles89} P. Carles and A. M. Cazabat, Colloids Surf. $\bf41$, 97 (1989).

\bibitem{Bahadur09} P. Bahadur, P. S. Yadav, K. Chaurasia, A. Leh, and R. Tadmor, J. Colloid Interf. Sci. $\bf332$, 455 (2009).

\bibitem{Sellier13} M. Sellier,  V. Nock, C. Gaubert, and C. Verdier, Eur. Phys. J. Special Topics $\bf219$, 131 (2013).

\bibitem{Cira15} N. J. Cira, A. Benusiglio, and  M. Prakash, Nature $\bf519$, 446 (2015).

\bibitem{Deegan00} R. D. Deegan, O. Bakajin, T. F. Dupont, G. Huber, S. R. Nagel, and T. A. Witten, Phys. Rev. E $\bf62$, 756 (2000).

\bibitem{Kobay10} M. Kobayashi,  M. Makino, T. Okuzono, and M. Doi, J. Phys. Soc. Jpn. $\bf79$, 044802 (2010).

\bibitem{Pradhan15} T. K. Pradhan and P. K. Panigrahi, Colloids Surf. A $\bf482$, 562 (2015).

\bibitem{Daniel01} S. Daniel, M. K. Chaudhury, and J. C. Chen, Science $\bf291$, 633 (2001).

\bibitem{Wasan01} D. T. Wasan, A. D. Nikolov, and H. Brenner, Science $\bf291$, 605 (2001).

\bibitem{Chaudhury92} M. K. Chaudhury and G. M. Whitesides, Science $\bf256$, 1539 (1992).

\bibitem{Brzoska93} J. B. Brzoska, F. Brochard, and F. Rondelez, Langmuir $\bf9$, 2220 (1993).

\bibitem{Gallardo99} B. S. Gallardo, V. K. Gupta, F. D. Eagerton, L. I. Jong, V. S. Craig, R. R. Shah, and N. L. Abbott, Science $\bf283$, 57 (1990).

\bibitem{Ichimura00} K. Ichimura, S. K. Oh, and M. Nakagawa, Science $\bf288$, 1624 (2000).

\bibitem{Style13}  R. W. Style, \emph{et al.} Proc. Natl Acad. Sci. USA $\bf110$, 12541 (2013).

\bibitem{Brochard89} F. Brochard, Langmuir $\bf5$, 432 (1989).

\bibitem{Xu12} X. P. Xu and T. Z. Qian, Phys. Rev. E $\bf85$, 051601 (2012).

\bibitem{Doi13} M. Doi, \emph{Soft Matter Physics (Oxford University Press, New York, 2013)}.

\bibitem{Doi15} M. Doi, Chin. Phys. B $\bf24$, 020505 (2015).

\bibitem{Xianmin16} X. M. Xu, Y. N. Di, and M. Doi, Phys. Fluids $\bf28$, 087101 (2016).

\bibitem{Man16} X. K. Man and M. Doi, Phys. Rev. Lett. $\bf116$, 066101 (2016).

\end{thebibliography}
\end{document}